\documentclass[runningheads]{llncs}
\usepackage[T1]{fontenc}
\usepackage{graphicx}
\usepackage{csquotes}
\usepackage{url}
\usepackage{soul}
\usepackage{tikz}
\usepackage{booktabs} 
\newlength{\myhspace}
\newcommand{\sevenPointScale}{\hspace{\myhspace} 1 \hspace{\myhspace} 2 \hspace{\myhspace} 3 \hspace{\myhspace} 4 \hspace{\myhspace} 5 \hspace{\myhspace} 6 \hspace{\myhspace} 7 \hspace{\myhspace}}
\begin{document}
\title{Digital Twins for Extended Reality Tourism: User Experience Evaluation Across User Groups \thanks{This work was supported by the European Union’s Horizon Research and Innovation Program under Grant 101092875 (DIDYMOS-XR: Digital DynaMic and responsible twinS for XR).}}
\titlerunning{DTs for XR Tourism: UX Evaluation Across User Groups}
%
\author{Maximilian Warsinke\inst{1}\orcidID{0009-0004-0264-5619} \and
Francesco Vona\inst{2}\orcidID{0000-0003-4558-4989} \and
Tanja Kojić\inst{1}\orcidID{0000-0002-8603-8979} \and
Jan-Niklas Voigt-Antons\inst{2}\orcidID{0000-0002-2786-9262} \and
Sebastian Möller\inst{1,3}\orcidID{0000-0003-3057-0760}}
\authorrunning{M. Warsinke et al.}
%
\institute{Quality and Usability Lab, Technische Universität Berlin, Berlin, Germany \and
Immersive Reality Lab, Hochschule Hamm-Lippstadt, Lippstadt, Germany \and
Deutsches Forschungszentrum für Künstliche Intelligenz, Berlin, Germany\\}
\maketitle              
\begin{abstract}
This study evaluates the user experience (UX) in extended reality (XR) tourism of two digital twin-based applications: an Augmented Reality Virtual Tour (AR-VT) for enhanced on-site visits and a Virtual Reality Virtual Tour (VR-VT) for remote exploration. Using a quantitative exploratory approach, 84 participants from Spain and Germany, divided into three sample groups, assessed UX, task load, presence, cybersickness, and emotional response through standardized questionnaires. Findings indicate that both applications provided a low task load and high enjoyment. The VR-based tour enhanced presence but posed usability and cybersickness challenges, while the AR-based tour achieved high UX ratings, with qualitative feedback suggesting areas for refinement. Correlation analysis revealed significant relationships between age, prior XR experience, and technological affinity with the measured metrics for both applications. These results highlight the importance of well-designed experiences tailored to XR novices, reinforcing the critical role of UX in digital twin-based XR tourism.

\newcommand\copyrighttext{%
  \footnotesize \textcopyright\ 2026 The Author(s), under exclusive license to Springer Nature Switzerland AG. This is the author’s accepted manuscript. The final version was presented at XR Salento 2025 and is published in Lecture Notes in Computer Science (LNCS, volume 15738). Available online at: \url{https://doi.org/10.1007/978-3-031-97769-5_3}

}

\newcommand\copyrightnotice{%
\begin{tikzpicture}[remember picture,overlay,shift={(current page.south)}]
  \node[anchor=south,yshift=10pt] at (0,0) {\fbox{\parbox{\dimexpr\textwidth-\fboxsep-\fboxrule\relax}{\copyrighttext}}};
\end{tikzpicture}%
}

\copyrightnotice

\end{abstract}
\setcounter{footnote}{0}
\section{Introduction}
Integrating digital twins (DTs) into tourism presents a promising opportunity to create novel extended reality (XR) experiences. Although the concept originated in an industrial context, DTs are now applied to various fields. Traditionally, a DT consists of a physical entity linked to a virtual counterpart through a data stream, enabling synchronization and interaction \cite{Jones_Snider_Nassehi_Yon_Hicks_2020}.

Recent DTs encompass systems, machines, buildings, and even entire cities, transmitting data unidirectionally or multidirectionally. While DTs often feature a 3D digital representation, they may also integrate sensor data (e.g., temperature and air quality) to enhance their utility. In the context of tourism, DTs can be created by reconstructing buildings and urban environments using video and depth capture techniques, forming a visual layer that serves as a foundation for immersive applications. XR technologies, such as augmented reality (AR) and virtual reality (VR), provide immersive and interactive 3D interfaces for engaging with DTs through head-mounted displays or smartphones.

Currently, tourism DTs are primarily used for cultural heritage sites that can be visited through VR. However, other use cases, such as destination management for stakeholders or experience enhancement through AR, are emerging \cite{de2025digital}. Demonstrators have been developed for remote visits to inaccessible sites and rural destinations \cite{sang2022} or commercial metaverse presences for businesses \cite{gallist23}. However, only a few user evaluations of XR DT applications have been conducted. This raises important questions about whether traditional XR user testing translates to DT-based applications and if the reconstructed 3D graphics can match conventional XR environments. This creates an opportunity to leverage UX validations for XR Tourism with DTs.

As part of the DIDYMOS-XR\footnote{\url{www.didymos-xr.eu}} project, DTs were developed for two contrasting cities: Vilanova i la Geltrú in Catalonia, Spain, and Etteln in North Rhine Westphalia, Germany. These DTs were integrated into two applications designed as use cases for XR tourism: Augmented Reality Virtual Tour (AR-VT) and Virtual Reality Virtual Tour (VR-VT). The AR-VT approach allows on-site exploration of the city DT with content overlaid at the user’s GPS location. The VR-VT application was developed for remote visits through immersive exploration of the city DT. At the project's current stage, the integrated DTs focus on the visual layer with unidirectional data, possibly extended with dynamically updated content in the future.

To evaluate these applications, a quantitative user study was conducted using an exploratory in-field approach with participant groups from Etteln and Vilanova i la Geltrú. The metrics selected were UX, task load, and emotion for AR-VT, and UX, presence, cybersickness, and emotion for VR-VT. Additionally, custom questionnaires were developed to receive ratings for use case-specific features, and open-ended questions for additional qualitative feedback. The participants recruited from Germany and Spain covered a wide age span and showed various levels of expertise in XR technology, forming a broad sample for evaluation.

The results indicated that both applications were generally well-received, with participants finding them enjoyable and the tasks manageable. However, usability issues and cybersickness were observed in some cases, particularly within certain sample groups. Correlation analysis between user factors and dependent variables revealed significant relationships, especially with affinity for technology interaction, age, and prior XR experience as influencing factors for the measured metrics.

The insights gained from this research will contribute to developing more user-friendly and impactful XR tourism experiences that use DTs and towards standardized evaluation methodologies for validating such applications.

\section{Related Work}
Various studies have explored the potential of DT technology in tourism by employing different evaluation methods to assess its impact on UX, competitiveness, data governance, and sustainability.

Deng et al. examined the role of DT design in metaverse tourism, emphasizing the importance of well-structured virtual environments for enhancing user engagement and influencing travel decisions \cite{Deng2024}. This research employs an experimental design with a between-subjects approach to analyze how different configurations shape tourists’ metaverse experiences. The findings indicate that specific DT elements, such as immersive features and interactive components, enhance user engagement and positively affect tourists’ physical travel intentions. 

A broad perspective is provided by a systematic literature review that synthesizes the existing research on DTs in tourism \cite{de2025digital}. The study categorizes implementations based on tourism type, spatial scale, and data integration, revealing that cultural tourism, particularly digitizing heritage sites, dominates the field. The review identifies a lack of synchronized physical-digital linkages. It highlights the early-stage nature of DT applications in tourism, focusing more on practical implementations than theoretical advancements.

Gallist explored the construction of a commercially viable DT as a demonstrator in an alpine city for metaverse tourism \cite{gallist23}. It details the process of creating an interactive virtual replica, including elements such as virtual shops, immersive streetscapes, and digital engagement tools, to encourage repeat visits and provide a scalable model for tourism businesses seeking to establish a metaverse presence. The trials in this case study indicated that users place greater value on functionalities within the application than on their visual quality alone. Another case study was conducted in research on rural cultural tourism, where a DT platform was developed using Unreal Engine \cite{sang2022}. This study explores how immersive virtual replicas of rural cultural sites can enhance tourism engagement and improve visitor experience.

The role of DTs in tourism competitiveness was investigated by Litaveniece et al., by focusing on the food service sector \cite{Litavniece2023}. A qualitative research approach was applied, incorporating pilot interviews with tourism industry representatives to evaluate key performance indicators. In another approach, model-based evaluation was applied to a smart tourism visual analysis platform that integrated real-time data on traffic, security, and environmental monitoring \cite{li24}. This study highlights the role of DTs in facilitating data-driven decision-making for tourism management.

Research on the intersection of DTs, big data governance, and sustainable tourism proposed a framework for evaluating compliance, transparency, and trust in DT applications \cite{Rahmadian2023}. A case study based on mobile positioning data examined how governance structures impact tourism management, ensuring regulatory adherence while enhancing data utilization. This research highlights the importance of integrating AI-driven analytics with DT systems to improve decision-making while addressing concerns related to privacy, security, and stakeholder communication. Another study on DTs in smart tourist destinations employed a mixed-method approach that combined expert opinions, real-world tourism data, and secondary sources \cite{florido24}. The findings emphasize that DTs are emerging as key tools for enhancing destination competitiveness and sustainability, offering governments and tourism organizations a data-driven approach to managing tourist flows and optimizing services.

The present work builds upon the methodology described in a previous publication \cite{vona2025}, which introduced a user-centric approach for evaluating DT applications across different use cases, including the two XR applications discussed in this study.

\section{Methods}

\subsection{Study Design}
The main objective was to evaluate the usability of the developed XR applications. This validation serves as an interim assessment at the midpoint of the project duration, informing the next iteration step of use cases and applications. The study followed an exploratory, quantitative usability testing approach, incorporating additional key UX-related measures in XR, such as cybersickness and presence, through standardized questionnaires.  Custom questionnaires were also designed to gather use case-specific feedback. Furthermore, the participants were engaged in conversations with the study supervisors for additional qualitative feedback. All validation procedures were approved by the ethics commission Lippstadt (approval numbers: EL202410151-6).

\subsection{Participants}
\subsubsection{Etteln Group}
The Etteln Group consisted of 16 participants, divided into three subgroups: elderly (n = 7), young (n = 5), and middle-aged (n = 4) individuals, resulting in a total of 16 participants. The mean age of the sample was 47.81 years (SD = 28.88), ranging from 16 to 79 years. The gender distribution was balanced, with eight males and eight females. Participants reported a mean experience with XR technologies of 2.5 (SD = 1.41) on a 7-point Likert scale. Prior to the study, seven participants had used VR via an HMD, and five had experienced VR through a smartphone. The mean ATI score was 3.79 (SD=0.99). Seven individuals from the group were retired, including one former teacher and one housewife. Three members indicated they were students. Two individuals were employed in the education sector, two worked in the administrative and corporate sectors, one worked as an IT consultant, and one was a carpenter. For clarity, this sample is referred to as the Etteln Group throughout this paper.

\subsubsection{Fair Group}
The Fair Group consisted of 50 participants with a mean age of 37.76 years (SD = 12.29) and an age range of 18 to 67 years. The group included 30 male and 20 female participants. Participants reported a mean XR experience of 3.54 (SD = 1.68) on a 7-point scale. Regarding prior experience with AR technology, 25 participants had used AR on smartphones, and 11 had tried AR headsets. The mean ATI score was 4.14 (SD = 0.82). Among the indicated professions, 11 individuals worked in education and research, including teachers and researchers. Nine participants were from the engineering and technology fields, including software engineers, metrologists, and VR developers. Five persons were in marketing and business management. The arts, media, and design sector was represented by four individuals, including photographers, designers, and artists. Four individuals were in administrative and corporate roles, such as banking and human resources. Another four worked in construction and industry. Two people were students, and one was in the healthcare domain. This group is referred to as the Fair Group in the following.

\subsubsection{Vilanova Group}
The Vilanova Group comprised 18 participants with a mean age of 40.72 years (SD = 12.67) and an age range of 20 to 62 years, including six males and 12 females. Their mean XR experience level was 3.61 (SD = 1.75). Regarding prior exposure to AR, 10 participants had used AR on smartphones, and 7 had experience with AR headsets. Their mean ATI score was 3.85 (SD = 0.80). Regarding the professions, seven individuals were working in the tourism and cultural heritage sector, including museum curators and professionals. The administrative and corporate sectors included five participants. The education and research field consisted of four persons, including teachers and librarians. Two worked in marketing and business management, and one individual in the video game industry. This group is called the Vilanova Group throughout the paper.

\subsection{Locations}
Validations were conducted in the field in Etteln and Vilanova i la Geltrú for multiple reasons. As the Etteln Group consisted of village inhabitants, visiting them promised a higher participation rate than inviting them to a larger city. Additionally, the AR-VT application was planned to be tested in front of the DT's physical counterpart, the community hall in Etteln. The Vilanova Group was also tested directly at the intended location around the \textit{Víctor Balaguer Museum}. The Fair Group was recruited ad hoc on the \textit{Fira de Novembre} and tested directly on-site.

\textbf{Etteln} is a village in North Rhine-Westphalia, Germany, with approximately 2,000 inhabitants. The village is involved in several projects related to smart cities and DT technology. The community hall in the center of the village was used to conduct user tests inside the building for the VR-VT and in front of the building for the AR-VT. A prominent point of interest (POI) that is part of the DT is the \textit{St. Simon und Judas Thaddäus} church, located 300 m from the community hall.

\textbf{Vilanova i la Geltrú} is a city in Catalonia, Spain, with approximately 68,000 inhabitants. The key POI for the AR-VT application is the \textit{Víctor Balaguer Museum}, where user tests were conducted in and around the building for the Vilanova Group. The Fair Group was recruited and tested at the \textit{Fira de Novembre}, one of the most important multi-sectoral outdoor exhibitions in Catalonia, featuring more than 190 exhibitors.

\subsection{Tools and Devices}
The \textbf{VR-VT} application was developed with the Unity Engine (2023 LTS) using the XR Interaction Toolkit. Cinemachine was used to implement the dynamic camera systems and Shader Graph for weather simulation. For the validation, participants were equipped with a Meta Quest Pro.

The \textbf{AR-VT} application was developed using Unity Engine (2022 LTS) using Unity’s Universal Render Pipeline (URP) to develop a custom render pipeline for the portal effect. The application supports two hardware platforms: Meta Quest 3 (OpenXR SDK) and mobile devices (Google ARCore). For this validation phase, the application was only tested on smartphones, specifically on Samsung S23 and Samsung S24 for the Etteln Group, and Samsung Galaxy and Realme 5 for the Fair and Vilanova Group.

\subsection{Applications}
The following application descriptions have been adapted from a previous publication detailing the validation plan for the use cases \cite{vona2025}. As the applications remained unchanged, the descriptions were paraphrased and extended for clarity and completeness.

\begin{figure}
    \centering
    \includegraphics[width=0.8\linewidth]{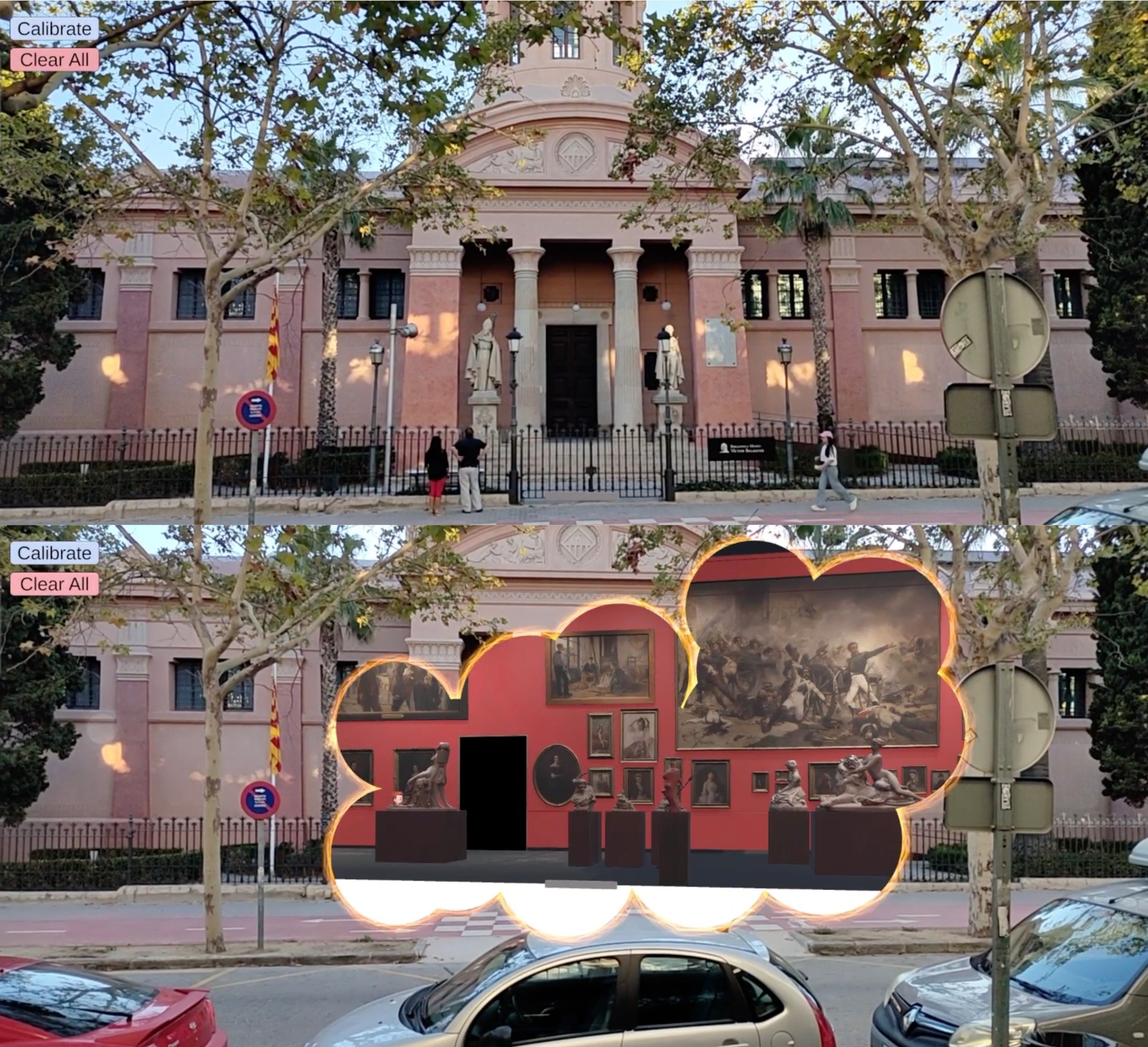}
    \caption{Two screenshots from the AR-VT application showing the initial view on the \textit{Víctor Balaguer Museum} in Vilanova i la Geltrú (top) and the DT view inside the museum through the virtual portal (bottom).}
    \label{fig:AR-VT_screenshot}
\end{figure}

The \textbf{AR-VT} application is a mobile application designed to allow the on-site exploration of city landmarks, even outside opening hours. Two versions were developed with content adapted to the city, displaying two different DTs of buildings for Etteln and Vilanova i la Geltrú. Users must be positioned in front of the target buildings where they can start the experience. Users can place a virtual AR portal by tapping it in the desired position. This portal serves as a gateway inside the building, allowing them to visually explore the building’s interior using AR, as depicted in Figure \ref{fig:AR-VT_screenshot}. In the Vilanova i la Geltrú version, users can examine the exhibits of the \textit{Víctor Balaguer Museum} from the sidewalk, whereas in Etteln, users can peek inside the community hall.

\begin{figure}
    \centering
    \includegraphics[width=0.8\linewidth]{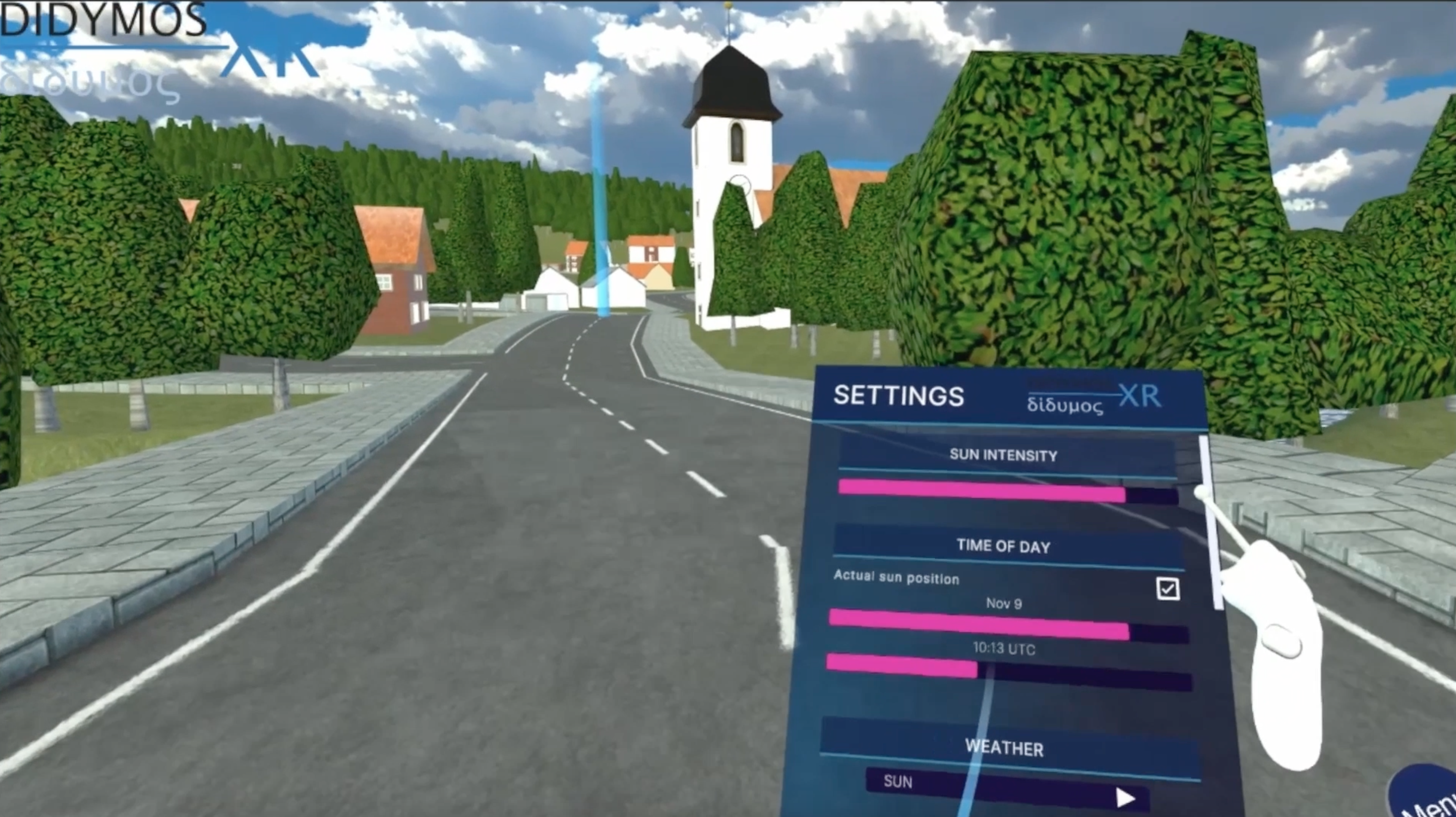}
    \caption{Screenshot of the VR-VT application on Ettelns DT main street with the \textit{St. Simon und Judas Thaddäus} church in the background and opened weather interface.}
    \label{fig:VR-VT_screenshot}
\end{figure}

The \textbf{VR-VT} application is an experience developed for Meta Quest, designed to provide an immersive remote exploration of Etteln. Users find themselves within the city DT in a first-person perspective. Using either continuous movement or teleportation, they can move between POIs that are highlighted through virtual light pillars. With their controllers, an interface can be opened that allows weather manipulation, shown in Figure \ref{fig:VR-VT_screenshot}. Different seasons and environmental conditions can be mimicked by adjusting the intensity of sunlight and by toggling rain. In addition, users can capture snapshots of the virtual environment using their controllers as cameras, simulating real-world photographs within the VR space to create virtual souvenirs.

\subsection{Questionnaires}
Starting with the \textbf{Demographics Questionnaire}, the participants were asked to complete their age, gender, profession, and prior experience with XR. Specifically, they were asked to self-report their experience with XR technology on a scale from 1 to 7 and to indicate whether they had already used VR on a headset, VR on a smartphone, AR on a headset, and AR on a Smartphone (cf. Appendix \ref{sec:demographics}). The \textbf{Affinity for Technology Interaction (ATI)} \cite{franke2019personal} is a 9-item questionnaire to assess the participants’ willingness and interest to engage in technological systems. A mean ATI score can be calculated from the 6-point Likert scales to determine the tech affinity of a participant or sample group. The short version of the \textbf{User Experience Questionnaire (UEQ-S)} \cite{schreppDesignEvaluationShort2017} with eight question items was used to measure the UX. From each of the four questions, ranging from -3 to 3, hedonic and pragmatic scores can be calculated, which can be averaged for the overall UX score. The \textbf{Self-Assessment Mannikin (SAM)}\cite{bradley1994measuring} is a pictorial questionnaire that measures the emotional dimensions of valence, arousal, and dominance. Each of the 5-point scales comes with a collection of images depicting a humanoid figure representing the emotional state. The \textbf{NASA Task Load Index (NASA-TLX)} \cite{Hart_Staveland_1988} was used only to validate the AR application. It measures the difficulty of completing a task within an application in terms of cognitive load and physical demand. Six question items were altered to a 7-point scale to fit the online questionnaire software instead of the originally designed continuous scale on paper. Instead of the NASA-TLX, the VR application was tested using two different questionnaires to cover the characteristics of the immersive aspects. The \textbf{iGroup Presence Questionnaire (IPQ)} \cite{igroup} was used to assess the feeling of being present in the virtual world. The 14-item questionnaire is divided into four dimensions: Overall presence, involvement, spatial presence, and perceived realism. All answers were given on a scale from 1 to 7. Additionally, the \textbf{Cybersickness in Virtual Reality Questionnaire (CSQ-VR)} \cite{CSQ-VR} was used to measure the appearance of cybersickness symptoms that may occur in VR. The questionnaire distinguishes between vestibular, oculomotor, and nausea symptoms, each rated on a scale from 1 to 7, complemented with an open-ended question to leave more specific symptom descriptions. Both the \textbf{Custom Questionnaires} for AR and VR included the overall enjoyment of the application and specific questions regarding the ease of use and usefulness of the implemented features. The question items can be found in Appendices \ref{sec:AR-VT} and \ref{sec:VR-VT}. The custom questionnaire was unintentionally not used for the Vilanova and Fair Group. For the participants in Etteln, the German versions of all standardized questionnaires were used, and the custom questions had to be translated because of insufficient language knowledge within the sample group. In Vilanova i la Geltrú, the participants used the English versions but gave some of their qualitative responses in Catalan, which were translated for evaluation.

\subsection{Procedures and Tasks}

For \textbf{AR-VT}, participants were given an overview of the project and the application to provide the context for the test, followed by an explanation of the tasks. Each participant received a pre-calibrated mobile device with the AR application. The user test began with the visualization of the augmented building guided by a portal hint. They opened the virtual portal, allowing them to transition to the interior of the digital building. They then interacted with an information pillar to access an event calendar. After completion, participants completed multiple questionnaires on a supplied laptop.

The \textbf{VR-VT} user test started with an overview of the project and application, as well as the tasks to be performed. Before starting the main test, participants underwent training to familiarize themselves with the VR headset and controls. The application was then launched, placing participants within the town's DT with the freedom to explore using two types of movement: free movement with the controller and teleportation to POIs. Participants also accessed the weather control system to modify conditions, such as lighting and rain, as well as a built-in screenshot functionality to capture DT images. Finally, the participants completed a questionnaire on a supplied laptop.

\subsection{Data Analysis}
Collected data were calculated according to the scoring of the standardized questionnaires and then analyzed using descriptive statistics in Python. Additionally, correlation analysis with the Spearman rank test was conducted using SciPy to examine the relationships between user factors and dependent variables. Comparative analyses between the sample groups were avoided due to numerous influencing factors, including variations in sample sizes and testing environments across different settings. Qualitative responses were collected, translated into English, and clustered according to reoccurring themes.

\section{Results}
\subsection{VR-VT}

\subsubsection{Presence}
The different dimensions of the IPQ received similar ratings, except for the perceived realism dimension. The general presence mean score was 3.94 (SD = 1.61). Involvement received a mean score of 4.16 (SD = 1.41), while spatial presence received a slightly higher score of 4.24 (SD = 0.85). In contrast, the perceived realism subscale received the lowest mean score of 2.41 (SD = 1.01).
\subsubsection{Cybersickness}
The results of the CSQ-VR indicated that the participants partially perceived cybersickness. The total score had a mean of 15.50 (SD = 7.82). Among the subscales, vestibular symptoms had the highest mean score of 5.81 (SD = 3.47), followed by oculomotor symptoms at 5.13 (SD = 2.70). Nausea symptoms had the lowest mean score of 4.56 (SD = 2.42). Participants explicitly mentioned eye strain, dizziness, and a fuzzy stomach in the complementary open-ended questions.
\subsubsection{User Experience}
The UEQ-S results indicated higher scores for hedonic quality than for pragmatic quality. The overall mean score was -0.10 (SD = 1.02). The hedonic dimension had a positive mean score of 0.19 (SD = 1.41), while the pragmatic dimension received a lower, negative mean score of -0.39 (SD = 1.19).
\subsubsection{Emotion}
The SAM scores indicated moderate emotional responses from the participants. The valence dimension had a mean score of 3.44 (SD = 0.89), whereas the arousal dimension received a mean score of 3.00 (SD = 0.89). The dominance dimension had the highest mean score of 3.31 (SD = 0.79).

\subsubsection{Custom Questionnaire}
The results of the custom questions provide insights into the specific functionalities of the application. Participants reported a high level of enjoyment, with a mean score of 4.44 (SD = 1.82). The ease of continuous navigation had a mean score of 4.00 (SD = 1.71) and the ease of teleportation movement was slightly higher at 4.13 (SD = 1.67). However, when asked about their preferred method of movement, only two participants indicated a preference for teleportation. The ease of reaching a POI received a mean score of 5.19 (SD = 1.52). Regarding environment settings, participants rated the change of weather feature with a mean score of 3.31 (SD = 2.02) and its usefulness at 3.44 (SD = 2.13). Finally, the ease of taking a snapshot had the lowest mean score of 2.94 (SD = 2.08), whereas its usefulness was rated higher at 3.69 (SD = 1.92). Qualitative responses highlighted difficulties with the snapshot function.

\subsubsection{Correlation Analysis}
\begin{figure}[ht]
    \centering
    \begin{minipage}{0.45\textwidth}
        \centering
        \includegraphics[width=\textwidth]{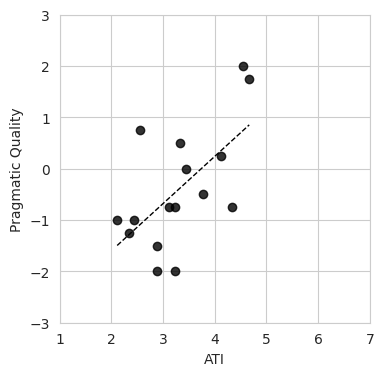}
    \end{minipage}
    \hfill
    \begin{minipage}{0.45\textwidth}
        \centering
        \includegraphics[width=\textwidth]{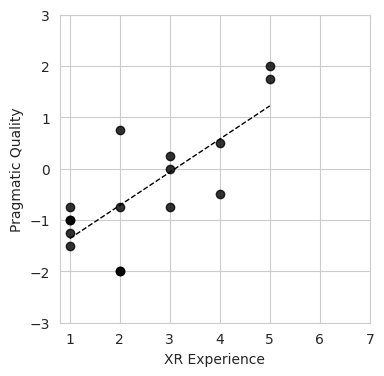} 
    \end{minipage}
    
    \vspace{0.2cm} 

    \begin{minipage}{0.45\textwidth}
        \centering
        \includegraphics[width=\textwidth]{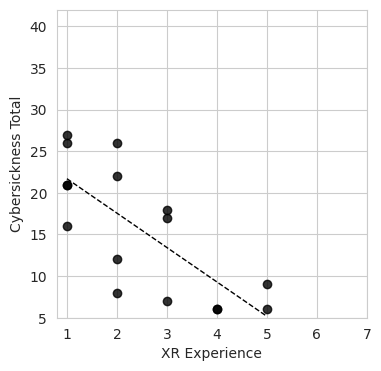}
    \end{minipage}
    \hfill
    \begin{minipage}{0.45\textwidth}
        \centering
        \includegraphics[width=\textwidth] {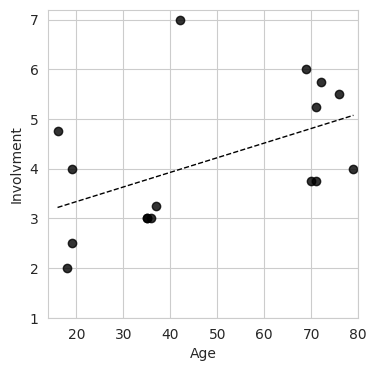} 
    \end{minipage}

\caption{Scatter plots depicting the relationships between ATI and pragmatic quality (top left), between XR experience and pragmatic quality (top right), XR experience and cybersickness total (bottom left) and age and involvement (bottom right) for the Etteln Group (N=16) in VR-VT. The dots represent individual data points, and the black regression lines indicate the correlations.}
\label{VR-VT_pragmatic}
\end{figure}

A correlation analysis was conducted between user factors and dependent measurements. Spearman’s rank-order correlation analysis revealed several significant relationships between the examined variables, partially visualized in Figure \ref{VR-VT_pragmatic}. The ATI scores ($\rho = 0.62$, $p < 0.01$) and prior experience with XR ($\rho = 0.73$, $p < 0.01$) exhibited a strong positive correlation with pragmatic quality from the UEQ-S. Additionally, prior experience with XR showed a strong negative correlation with the total cybersickness score measured by the CSQ-VR ($\rho = -0.74$, $p < 0.01$) and a moderate negative correlation with arousal from the SAM  ($\rho = -0.54$, $p < 0.05$). Age demonstrated a strong positive correlation with both the perceived realism ($\rho = 0.55$, $p < 0.05$) and involvement ($\rho = 0.53$, $p < 0.05$) dimensions of the IPQ.

\subsection{AR-VT}

\subsubsection{User Experience}
The UEQ-S results for the Etteln Group indicated varying deviations from the benchmark. The overall mean score was 0.70 (SD = 1.16), categorized as 'below average'. The mean pragmatic quality score of 0.65 (SD = 1.24) was classified as 'bad', while the mean hedonic quality score of 0.75 (SD = 1.49) fell into the 'above average' category. The Fair Group showed the most promising results among the groups. The overall mean score was 1.41 (SD = 1.12), with a mean hedonic score of 1.46 (SD = 1.29), and a mean pragmatic quality score of 1.36 (SD = 1.19). Both the overall and hedonic quality scores were classified as 'good', while the pragmatic quality score was categorized as 'above average'. The Vilanova Group gave lower ratings, with all three scores falling into the 'below average'. The overall mean score was 1.11 (SD = 1.11), with the mean hedonic quality score at 1.06 (SD = 1.00) and the mean pragmatic quality score at 1.17 (SD = 1.40).

\subsubsection{Task Load}
The NASA-TLX scores indicated a relatively low level of perceived workload in all the groups. The Etteln Group received an overall mean score of 2.72 (SD = 0.92). The Fair Group reported a mean score of 2.81 (SD = 1.07), while the Vilanova Group scored a slightly lower mean score of 2.5 (SD = 0.83).

\subsubsection{Emotion}
The SAM results showed similar ratings with varying degrees of intensity for the three groups. The Etteln Group reported a mean valence score of 4.00 (SD = 0.82), a mean arousal score of 2.00 (SD = 0.89), and a mean dominance score of 3.38 (SD = 0.71). Participants of the Vilanova Group reported a mean valence score of 3.89 (SD = 0.90), a mean arousal score of 2.33 (SD = 0.97), and a mean dominance score of 3.38 (SD = 0.98). The Fair Group reported a mean valence score of 4.30 (SD = 0.83), a mean arousal score of 1.66 (SD = 0.75), and a mean dominance score of 3.34 (SD = 0.72).

\subsubsection{Custom Questionnaire}
The custom questionnaire was only employed for the Etteln Group and underlined the satisfaction and ease of use of the application. Participants rated their overall enjoyment with a mean score of 5.37 (SD = 1.36). Regarding finding the location to place the frame, the mean ease of use was 4.56 (SD=1.31). Finding the POI and opening the portal received an easiness mean score of 6 (SD=1.41). Looking through the portal was rated as the easiest, with a mean of 6.19 (SD=1.42). Additionally, when asked to choose between devices, 4 out of 16 users (25 \%) indicated a preference for using an AR headset, such as the HoloLens 2.

\subsubsection{Qualitative Feedback}
Participants in the Fair Group provided qualitative feedback on their experiences during the debriefing, highlighting concerns related to the usability of the application. Users pronounced the wish to move inside the virtual augmented environment or zoom to examine the objects more closely. Additionally, the application's light rendering was insufficient, impacting the visibility of details in artworks, and the image sharpness was inconsistent. Informational content was mentioned multiple times, as users found a lack of descriptive or educational text accompanying the painting and sculptures. Overall, the feedback suggests that improvements in navigation, visual clarity, and informational content would enhance the experience.

\subsubsection{Correlation Analysis}
\begin{figure}[ht]
    \centering
    \begin{minipage}{0.45\textwidth}
        \centering
        \includegraphics[width=\textwidth]{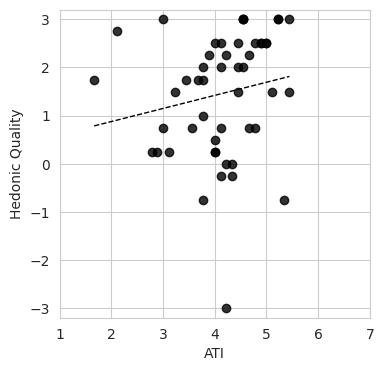}
    \end{minipage}
    \hfill
    \begin{minipage}{0.45\textwidth}
        \centering
        \includegraphics[width=\textwidth]{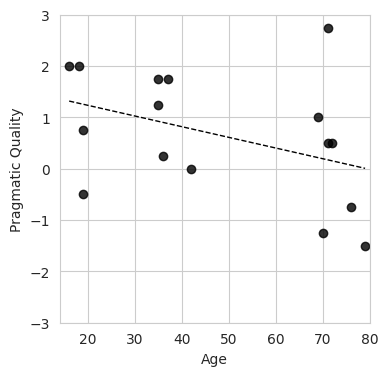} 
    \end{minipage}

\caption{Scatter plots showing the relationships between ATI and hedonic quality (left) for the Fair Group (N=50) in AR-VT and between age and pragmatic quality (right) for the Etteln Group (N=16) in AR-VT. The dots represent individual data points, and the black regression lines indicate the correlations.}
\label{fig2}
\end{figure}

Spearman’s rank correlation analysis was conducted to examine the relationship between user factors and dependent variables. Some of the significant correlations are displayed in Figure \ref{fig2}. In the Etteln Group, a moderate negative correlation was found between participants’ age and pragmatic quality as measured by the UEQ-S ($\rho = -0.50$, $p < 0.05$). In the Fair Group, a weak to moderate positive correlation was observed between ATI and overall UX ($\rho = 0.29$, $p < 0.05$) as well as between ATI and hedonic quality ($\rho = 0.31$, $p < 0.05$) and valence ($\rho = 0.29$, $p < 0.05$).
The Vilanova Group exhibited only one significant correlation, namely a negative moderate correlation of the participants’ age with dominance ($\rho = -0.52$, $p < 0.05$).

\section{Discussion}
\subsection{VR-VT}
The VR-VT validation results show mixed responses towards the application. Participants seemed to enjoy the experience, as indicated by the custom questionnaire results, and had a tendency towards a positive valence. However, the UX scores can be classified as "bad" according to the UEQ benchmarks. Specifically, the pragmatic dimension resulted in low scores, indicating insufficient usability of the application. Notably, these scores correlate with the participants’ prior XR experiences, indicating that novice VR users had more difficulties. Cybersickness symptoms occurred for some participants, especially those who were less experienced, which aligns with the findings of Doty et al., who recently confirmed that repeated exposure to VR can reduce cybersickness \cite{doty24}. From the IPQ scores, it can be assumed that the participants felt present in the virtual environment, but that the fidelity of the DT may not be high enough for a convincing effect of perceived realism. This might be due to the approach of the DTs 3D reconstruction, which may currently not fulfill the expectations of users. Interestingly, perceived realism and involvement scores correlate with the participants’ age, possibly because younger participants are more exposed to high-quality computer graphics in their daily life, as video games are popular among the younger generation, creating higher expectations for graphics quality. While the sample size here is limited, the findings align with those of Dilanchian et al. \cite{Dilanchian_Andringa_Boot_2021}, who also found higher IPQ scores for older adults, following the same argumentation of less exposure.

\subsection{AR-VT}
The Etteln version of the AR-VT application received the worst UX scores by its sample group. The hedonic quality was rated better than pragmatic quality, indicating problems with the usability of the application. However, the task itself was not too demanding, as shown by the task load results. Despite diminished usability, the Etteln Group participants maintained a positive valence and felt in control. The custom questionnaire underscored the enjoyment and ease of use of these functionalities. Interestingly, the correlation analysis revealed that age and pragmatic quality were negatively correlated, meaning that older participants found the application less usable than younger participants. While the Etteln Group tested a different version of the application, thus preventing a direct comparison to the Vilanova and Fair Group, it can be noted that the mean XR experience, as indicated in the demographics questionnaire, was the lowest out of the three groups, possibly affecting the overall experience negatively.

The Vilanova i la Geltrú version of the application was perceived slightly differently by the Fair and Vilanova Group. The Fair Group showed the best UX with a slight favor of hedonic over pragmatic qualities. In contrast, the Vilanova group rated the UX only as below average, with pragmatic quality slightly higher than hedonic quality. The task load was rated as low and similar in both the groups. Regarding valence, the participants seemed to have a positive and not arousing experience with a moderate degree of control, this is true for both groups, with more extreme values in the fair group. However, multiple improvements have been articulated through qualitative feedback. Specifically, the lack of movement in the virtual world has been criticized. In addition, the visual quality can be improved in terms of sharpness and light rendering. Correlation analysis revealed technological affinity as the most influential independent factor for the Fair Group, where higher ATI scores can be associated with improved UX, especially with hedonic quality and valence. This leads to the assumption that technological affinity affects the experience positively. Possibly, the motivation to engage with the technical system helped overcome usability obstacles.

\section{Conclusion}
The integration of DTs as a foundation for XR tourism applications presents a high potential for enjoyable experiences. The results from user testing of the AR-VT and the VR-VT applications indicate that while these technologies provide engaging experiences through feasible tasks, several challenges remain in optimizing usability and visual fidelity. The evaluations revealed relationships in the perception of the applications with age, prior XR experience, and technological affinity. Higher age was associated with a higher sense of involvement and perceived realism and lower pragmatic quality ratings. XR experience was linked to lower cybersickness and higher pragmatic quality, while high ATI scores relate to a higher overall UX score. These insights suggest that while the DT-based applications developed are promising, future iterations must focus on improving graphical fidelity, reducing usability issues, and mitigating side effects, such as cybersickness, particularly for novice XR users. Moreover, the custom feedback points to the need for more dynamic interaction features. As XR and DT technologies evolve, these findings provide valuable guidance for future developments in the field, ensuring that DT-based experiences can reach their full potential for creating immersive and user-friendly tourist solutions.

\subsection{Future Work}
Future research should conduct comparative studies to further investigate the relationships identified in this study, particularly the interactions between technology affinity, age, and XR experience. Understanding how these factors influence each other will be crucial for designing broadly accessible XR applications.
Additionally, controlled validation settings should be implemented to enable direct group comparisons, isolating specific influencing factors. Evaluation methodologies for XR applications should also be expanded to include DT-specific metrics, ensuring the continued advancement of this emerging field.

\subsection{Limitations}
Although validations were conducted in the field, not all tests took place in the intended end-user locations. For example, the AR experience was conceptualized to be used in front of the \textit{Víctor Balaguer Museum}, which makes the findings from the Fair Group not directly transferrable to the sidewalk scenario. Similarly, the VR-VT application was tested in the community hall with Etteln residents, meaning the findings may not fully reflect the experience of tourists unfamiliar with the village. Another limitation was sample size.  Originally, the aim for participant count was a minimum of 20 users. For two of the groups, this requirement was not met, reducing the generalizability of the findings. A factor that was left out of discussion was possible cultural differences in evaluation between participants from Germany and Spain.

\subsection*{Acknowledgements}
The AR-VT application was developed under the leadership of the University of Patras, whereas the VR-VT application was led by the Digital Twin Technology GmbH. User testing was conducted by FIWARE Foundation in Etteln and Neàpolis in Vilanova i la Geltrú. During the preparation of this work, the authors used GPT-4 and 'Paperpal for Overleaf' in order to: Grammar and spelling check. After using these tools, the authors reviewed and edited the content as needed and take full responsibility for the publication’s content.

%
%
%
\bibliographystyle{splncs04}
\bibliography{main}

\appendix
\section{Appendix - Custom Questionnaires}
\subsection{Demographics (English Version)}
\label{sec:demographics}
\begin{enumerate}
    \item Gender
    \begin{itemize}
        \item Female
        \item Male
        \item Other
    \end{itemize}
    \item Age:
    \item Profession:
    \item Are you a Student?
    \begin{itemize}
        \item Yes
        \item No
    \end{itemize}
    \item How experienced are you with Extended Reality technologies? (Virtual Reality, Augmented Reality, Mixed Reality)
\begin{itemize}
\item Not all all \sevenPointScale Very experienced
\end{itemize}
    \item What of the following technologies have you used before?
    \begin{itemize}
        \item Virtual Reality on Headset
        \item Virtual Reality on Smartphone
        \item Augmented Reality on Headset
        \item Augmented Reality on Smartphone

    \end{itemize}

\end{enumerate}

\subsection{AR-VT (English Version)}
\label{sec:AR-VT}
\begin{enumerate}
    \item How much did you enjoy the experience?
\begin{itemize}
    \item Not at all \sevenPointScale Enjoyed very much
\end{itemize}
    \item How easy was it to find the place to frame?
\begin{itemize}
    \item Very difficult \sevenPointScale Very easy
\end{itemize}
    \item How easy was it to find the point of interest?
\begin{itemize}
\item Very difficult \sevenPointScale Very easy
\end{itemize}
    \item How easy was it to open the portal?
\begin{itemize}
\item Very difficult \sevenPointScale Very easy
\end{itemize}
    \item How easy was it to look inside the portal?
\begin{itemize}
\item Very difficult \sevenPointScale Very easy
\end{itemize}
\item Would you recommend other gestures to open the portals? 
\begin{itemize}
    \item Yes
    \item No
\end{itemize}
\item If yes, which one?
\item Do you think a headset like the one in the figure (HoloLens 2) would have been better for this experience instead of the smartphone?
\begin{itemize}
    \item Yes
    \item No
\end{itemize}
\item Additional comments:
\end{enumerate}
\subsection{VR-VT (English Version)}
\label{sec:VR-VT}
\begin{enumerate}
    \item How much did you enjoy the experience?
\begin{itemize}
\item Not at all \sevenPointScale Enjoyed very much
\end{itemize}
    \item How easy was it to navigate with the free movement?
\begin{itemize}
\item Very difficult \sevenPointScale Very easy
\end{itemize}
    \item How easy was it to navigate with teleportation?
\begin{itemize}
\item Very difficult \sevenPointScale Very easy
\end{itemize}
    \item How easy was it to reach a point of interest?
\begin{itemize}
\item Very difficult \sevenPointScale Very easy
\end{itemize}
    \item What movement did you prefer?
\begin{itemize}
    \item Free Movement
    \item Teleportation
\end{itemize}
    \item How easy was it to take a snapshot?
\begin{itemize}
\item Very difficult \sevenPointScale Very easy
\end{itemize}
    \item How would you rate the usefulness of this feature?
\begin{itemize}
\item Not at all \sevenPointScale Very useful
\end{itemize}
    \item Additional comments:
\end{enumerate}

\end{document}